\documentclass[hyper]{JHEP3}
\pdfoutput=1

\usepackage{epsfig}
\usepackage{latexsym}
\usepackage{amsfonts}
\usepackage{amsmath}
\usepackage{amsthm}
\usepackage{amssymb}
\usepackage{amsbsy}

\newcommand{\C}[1]{C^{(#1)}}

\newcommand{\ul}[2]{^{#1}{}_{#2}}

\newcommand{\transp}[1]{{#1}^{\mathrm{\textit{\tiny{T}}}}}

\def\calf         {{\cal F}}
\def\calg         {{\cal G}}

\def\calh         {{\cal H}}
\def\cali         {{\cal I}}

\def\calk         {{\cal K}}

\def\calm         {{\cal M}}
\def\caln         {{\cal N}}

\def\calp         {{\cal P}}

\def\cals         {{\cal S}}

\def\calz         {{\cal Z}}

\def\rez          {{\re\left[e^{-i\a}\vec{\calz}\right]}}
\def\imz		  {{\im\left[e^{-i\a}\vec{\calz}\right]}}

\newsavebox{\uuunit}
\sbox{\uuunit}
    {\setlength{\unitlength}{0.825em}
     \begin{picture}(0.6,0.7)
        \thinlines
        \put(0,0){\line(1,0){0.5}}
        \put(0.15,0){\line(0,1){0.7}}
        \put(0.35,0){\line(0,1){0.8}}
       \multiput(0.3,0.8)(-0.04,-0.02){12}{\rule{0.5pt}{0.5pt}}
     \end {picture}}

\def\be{\begin{equation}}
\def\ee{\end{equation}}
\def\bea{\begin{eqnarray}}
\def\eea{\end{eqnarray}}

\def\a{\alpha}

\def\l{\lambda}

\def\m{\mu}
\def\n{\nu}

\def\r{\rho}

\def\s{\sigma}

\def\re{\mathrm{Re}}
\def\im{\mathrm{Im}}
\def\Tr{\mathrm{Tr}}
\def\rot{\vec{\nabla}\times}
\def\grad{\vec{\nabla}}

\title{BPS dyons and Hesse flow}
\author{Dieter Van den Bleeken\footnote{Email: dieter.vdbl@gmail.com}\\

Galatasaray University,\\
Department of Mathematics,\\
34357 Ortak\"{o}y-Istanbul, Turkey}

\abstract{We revisit BPS solutions to classical $\caln=2$ low energy effective gauge theories. It is shown that the BPS equations can be solved in full generality by the introduction of a Hesse potential, a symplectic analog of the holomorphic prepotential. We explain how for non-spherically symmetric, non-mutually local solutions, the notion of attractor flow generalizes to gradient flow with respect to the Hesse potential. Furthermore we show that in general there is a non-trivial magnetic complement to this flow equation that is sourced by the momentum current in the solution.}

\begin{document}
\section{Introduction and summary}\label{secintro}
Theories with $\caln=2$ supersymmetry in 4 dimensions seem to strike a unique balance between analytical control and non-trivial physics. This is most famously illustrated by the work of Seiberg and Witten \cite{Seiberg:1994rs, Seiberg:1994aj} in gauge theory and that of Maldacena, Strominger and Witten \cite{Maldacena:1997de} in the presence of gravity. What makes $\caln=2$ theories so special and suitable to an exact analysis is the presence of BPS states. Finding the spectrum of stable states of a given physical theory is in general a formidable problem, especially at strong coupling. The simpler problem of computing the BPS spectrum remains very challenging, but recently a number of new insights have led to tremendous progress that suggests a complete and exact solution to this and other problems in $\caln=2$ theories might be within reach. See e.g. \cite{Denef:2007vg, deBoer:2008zn, Kontsevich:2009xt, Gaiotto:2009hg, Gaiotto:2009we, Dimofte:2009tm, Alday:2009aq, Manschot:2010qz, Cecotti:2011rv, Alim:2011ae, Kim:2011sc} for a small selection out of the everyday growing literature.

BPS states have been studied through a wide variety of methods. Among others one can think of them as non-Abelian dyons \cite{Weinberg:2006rq}, string-webs \cite{Bergman:1997yw}, WKB curves \cite{Gaiotto:2009hg}, attractors \cite{Ferrara:1995ih,Denef:2000nb} or ground states in (quiver) quantum mechanics \cite{Denef:2002ru,Alim:2011ae,Kim:2011sc}. Each of these representations has its merits and limitations. 

A particularly interesting regime is that of low energy. Studying BPS states in this limit has many advantages. Through electromagnetic duality we have an exact description at each value of the coupling \cite{Seiberg:1994rs}. Furthermore all gauge interactions are Abelian and this simplifies the theory a lot. Finally this description is directly formulated in physical 4d space-time, making it very intuitive. The formalism has mainly been developed in supergravity \cite{Denef:2000nb, Bates:2003vx, Denef:2007vg}, since there the UV (string) theory is only partially understood. This study of BPS solutions to $\caln=2$ supergravity led to an interesting conjecture. Denef proposed that BPS states and their stability are encoded in so called {\it split attractor flows} \cite{Denef:2000nb, Denef:2007vg}. To prove this conjecture we would need a much better and more detailed understanding of Type II string theory compactified on a Calabi-Yau manifold.

In the case of $\caln=2$ gauge theories the UV theory is under much better control and in certain cases the BPS spectrum is fully understood, in one or more of the descriptions mentioned above. This suggests that an analog of the split attractor flow conjecture in gauge theory could be subjected to precise tests. This could furthermore lead to a precise formulation and possibly a proof,  or if not, at least an understanding of its limitations. This paper can be seen as a first step towards formulating split attractor flows in $\caln=2$ gauge theory.

BPS solutions to low energy $\caln=2$ gauge theory have been considered in the literature before \cite{Chalmers:1996ya, Kol:2000tw, Argyres:2001bp, Popescu:2001rf,Lee:2011ph}. However, none of these have made a completely general analysis. Either one concentrated on some specific (simple) theories or the focus was on spherically symmetric solutions to an arbitrary $\caln=2$ theory. This last assumption is especially stringent, as it is well known that much of the spectrum manifests itself as solutions that consist of multiple centers carrying mutually non-local charges and breaking spherical symmetry \cite{Lee:1998nv, Denef:2000nb}.

In this paper we present the general solution to the BPS equations of any low energy $\caln=2$ gauge theory without matter. In section \ref{secle} we will review some essential properties of $\caln=2$ gauge theories and their low energy description in terms of $r$ massless vector multiplets and the prepotential $\calf$. We then write down the BPS equations in section \ref{secBPSsol}, and show how a general solution is determined in terms of $2r$ real harmonic functions\footnote{The precise description of the sources for these harmonic function and the interactions among them is an interesting problem that we shortly discuss in section \ref{comso} but of which we leave a general detailed treatment to future work.}. The relation to the original $r$ vector multiplets is shown to involve a Legendre transform of the imaginary part of the prepotential that is known in the supergravity literature as the Hesse potential $\cals$ \cite{Cardoso:2006bg, Mohaupt:2009iq,Cardoso:2010gc}. 

This Hesse potential first appeared in discussions of the special geometry of the theory \cite{Freed:1997dp, Hitchin:2005uu, Cortes:2001qd, Alekseevsky:1999ts, Ferrara:2006js, Ferrara:2006at}. It takes its name from the fact that in Darboux coordinates the metric on moduli space is nothing but the Hessian matrix of $\cals$. We review these facts in the first part of section \ref{dualandspecial} and appendix \ref{appmetr}. Since both the Hesse potential and the Darboux coordinates $Y^a$ give a duality covariant description of the special geometry we can use them to give a duality covariant description of the complete theory, including the U(1) gaugefields. We formulate the duality covariant Lagrangian (\ref{dualcovlag}) and its variation principle in the second part of section \ref{dualandspecial}. In appendix \ref{appdual} we show in detail the equivalence between this formulation and the standard one. In this section \ref{dualandspecial} we also rederive the BPS equations in a manifestly duality invariant form and at the end comment on the description of sources and the constraints on their positions.

In section \ref{genflow} we show how the attractor equations for the central charge, well established for spherically symmetric solutions, generalize to gradient flow equations determined by a central charge vector field. It is here that the Hesse potential finds its physical interpretation: evaluated on the solution it is the potential for this gradient flow, that we therefore refer to as Hesse flow (\ref{Hessflow}). Finally we show how the BPS equations imply a rather remarkable set of equations for the central charge vector field. If we interpret its real part as an 'electric' field and its imaginary part as a 'magnetic' field then it satisfies Maxwell's equations (\ref{deMax1}, \ref{deMax2}). What is particularly intriguing is that the equations are sourced by the energy-momentum 4-current. In particular these equations imply that the Hesse potential evaluated on the solution equals the gravitational potential of that solution (i.e. it solves the Poisson equation sourced by the energy density). Furthermore in cases without spherical symmetry and mutually non-local charges there is an additional magnetic equation sourced by the momentum current present in the solution.

\section{$\caln=2$ gauge theory at low energy}\label{secle}
In this section we shortly review some notions in $\caln=2$ gauge theory and in particular their low energy description. We will only consider results that are relevant to the rest of the paper and use this section mainly to fix our notation.

In this paper we will restrict ourselves to the vector multiplet sector of $\caln=2$ rigid supersymmetry in 4 dimensions. We expect our results to remain essentially unchanged,  at least qualitatively,  when hypermultiplets and/or the gravity multiplet are included and some crucial notions are appropriately adjusted to those cases. 

\subsection{UV origin}
The $\caln=2$ vector multiplet consists of a gauge field $A_\mu$, a complex scalar $\phi$ and a Dirac spinor $\psi$, all valued in the adjoint representation of some gauge group G. The extended supersymmetry forbids the presence of a superpotential and so the scalar potential arises purely from integrating out auxiliary superfields and is given by
\be
V\sim \Tr[\phi,\phi^\dagger]^2\,.\label{scalpot}
\ee
In a generic vacuum the gauge group G will be spontaneously broken to its maximal torus U$(1)^r$, with $r\equiv\mathrm{rank}$ G. Furthermore, since the flat directions of the potential (\ref{scalpot}) are in one to one correspondence with generators of the Cartan subalgebra of G, there are also precisely $r$ massless scalar fields in such a vacuum. This implies that at low enough energies the theory is effectively described by the dynamics of $r$ massless $U(1)$ vector multiplets $(A_\mu^A,\phi^A,\psi^A)$.

$\caln=2$ supersymmetry is restrictive enough to fully determine the two-derivative low energy effective action up to a single holomorphic function, the prepotential $\calf$. In $\caln=2$ superspace the action takes the simple form
\be
S=\im\int\,d^4x\,d^4\theta\, \calf(W)\,.\label{supspaceact}
\ee
The precise form of $\calf$ depends on the details of the UV theory and encodes the effect of integrating out the various massive fields. Using arguments based on electromagnetic duality, Seiberg and Witten \cite{Seiberg:1994rs} provided a geometric way to efficiently and exactly compute $\calf$ in the case G$=$SU(2). This result has since been generalized to include a wide variety of gauge groups. Some of the state of the art can be found in \cite{Tachikawa:2011yr}, together with a nice overview of different techniques and further references to the literature.

In this paper we will make no further assumptions about $\calf$ and start directly from the low energy effective action, leaving the rank $r$ and prepotential $\calf$ as arbitrary input parameters. Readers interested in more concrete applications can directly apply our formalism to any specific $r$ and $\calf$ obtained from a given UV theory.

\subsection{Low energy bosonic Lagrangian}
Let us now take a closer look at these low energy $\caln=2$ effective actions. The bosonic action following from (\ref{supspaceact}) is
\be
S=-\int d^4x\,\im\left[\tau_{AB}\left(\frac{1}{4}F^A_{\m\n}(F^{B\,\m\n}+i\star\!F^{B\,\m\n})+\frac{1}{2}\partial_\m\phi^A\partial^\m\bar\phi^B\right)\right]\,.\label{bosac}
\ee
As explained above, the fields in this Lagrangian are $r$ complex scalars $\phi^A$ and $r$ U(1) gaugefields $A_\mu^A$, labeled by an index $\mbox{\footnotesize{$A$}}=1,\ldots,r$.
All couplings in the Lagrangian (\ref{bosac}) depend on the scalar fields and are determined in terms of the prepotential $\calf(\phi)$. The kinetic term of the scalars is given by a non-linear sigma model with metric
\be
G_{AB}(\phi,\bar\phi)=\im\left[\tau_{AB}(\phi)\right]\,\quad\qquad\qquad\quad \tau_{AB}\equiv\partial_A\partial_B\calf\equiv\frac{\partial^2\calf}{\partial\phi^A\partial\phi^B}\,. \label{scmetr}
\ee 
A metric determined in such a way in terms of a holomorphic prepotential is called (rigid) special K\"ahler \cite{Craps:1997gp}. Its K\"ahler potential is of the special form 
\be
\calk=\im\left[\partial_A\calf\bar\phi^A\right]\,.\label{Kahlerpot}
\ee

The metric $G_{AB}$ also appears as a scalar dependent gauge coupling in front of the kinetic term of the U(1) gaugefields in (\ref{bosac}). Furthermore there is the term involving the dual fields $\star F^{A}$, note that in our notation
\be
\star F^{A\,\m\n}\equiv\frac{1}{2}\epsilon^{\m\n\r\s}F^{A}_{\r\s}\,.
\ee 
This term couples to the scalars through 
\be
\Theta_{AB}\equiv\re\left[\tau_{AB}\right]\,.
\ee 
The theory thus also contains a matrix of scalar dependent $\theta$-angles $\Theta_{AB}$. In summary, the matrix $\tau_{AB}=\Theta_{AB}+i G_{AB}$ plays the role of the complexified gauge couplings.

\subsection{Duality structure}\label{subsecdualstr}
The electromagnetic duality of Maxwell's equations generalizes to theories of the form (\ref{bosac}), where it is part of a larger duality group \cite{Gaillard:1981rj}. To make this duality manifest it is convenient to formally extend the U(1) field strengths $F^A$ to a doublet
\be
F^a\equiv \begin{pmatrix}F_A\\F^A\end{pmatrix}\,,\qquad\qquad F_A\equiv\re\left[\tau_{AB}(F^B+i\star F^B)\right]\,.\label{doubdef}
\ee
Here the index $a=1,\ldots,2r$ runs over both upper and lower indices $\mbox{\footnotesize{$A$}}=1,\ldots,r$. To create the doublet we introduced a new component $F_A$, that is defined in such a way that the Bianchi identity together with the equations of motion following from (\ref{bosac})  take a simple and elegant form in doublet notation: 
\be 
dF^a=0\,.
\ee
This form of the equations is manifestly invariant under GL$(2r,\mathbb{R})$ transformations. One should however keep in mind that $F^a$ can't take values in all of the abstract $2r$ dimensional vector space, since there is a relation between the top and bottom components. We can express this relation (\ref{doubdef}) in a covariant way as
\be 
\star F^a=\cali^a{}_b F^b\qquad\mbox{with}\qquad \cali^a{}_b=\begin{pmatrix}-\left[\Theta G^{-1}\right]_{A}{}^{B}&\ &G_{AB}+\left[\Theta G^{-1}\Theta\right]_{AB} \\ -\left[G^{-1}\right]^{AB}&\ &\left[G^{-1}\Theta\right]^{A}{}_{B}\end{pmatrix}\,.\label{self-dual}
\ee
One can check that $\cali$ is a complex structure, $\cali^2=-1$, so the above constraint is a generalized imaginary self-duality condition. Invariance under duality thus also imposes that this constraint remains invariant, i.e. $\star F^{\prime a}=\cali^{\prime a}{}_b F^{\prime b}$. Some algebra reveals that this is only the case for transformations in the subgroup Sp($2r,\mathbb{R}$)$\subset$GL$(2r,\mathbb{R})$ \cite{Gaillard:1981rj}.

Note that although the field strengths $F^a$ transform in the fundamental representation, the couplings $\tau_{AB}$ transform under the natural action of  Sp($2r,\mathbb{R}$) on the Siegel upper half-space. Split the symplectic matrix $M\in$ Sp($2r,\mathbb{R}$) into doublet notation as
\be
M=\begin{pmatrix}A&B\\C&D\end{pmatrix}\quad\mbox{with}\quad\transp{A}C-\transp{C}A=0\,,\qquad \transp{B}D-\transp{D}B=0\quad\mbox{and}\quad \transp{A}D-\transp{C}B=1\,.\nonumber
\ee
Then the couplings transform as follows under application of $M$:
\be
\tau_{AB}^\prime=\left[(A\tau+B)(C\tau+D)^{-1}\right]_{AB}\,.\label{couptransf}
\ee
These couplings are however not independent fields but determined in terms of the scalar fields as the second derivative of the prepotential: $\tau_{AB}=\partial_A\partial_B\calf$. To make their transformation consistent with this definition one introduces a dual scalar
\be
\phi_A\equiv \partial_A\calf\,.
\ee
One can now check that the transformation (\ref{couptransf}) follows if we assume that the scalars $\phi^A$ transform together with the $\phi_A$ in a fundamental doublet of the Sp($2r,\mathbb{R}$) duality group:
\be
\phi^a\equiv\begin{pmatrix}\phi_A\\\phi^A\end{pmatrix}\,,\qquad\quad \phi^{\prime a}=M^a{}_b \phi^b\ . \label{scalartransf}
\ee
Because duality mixes the scalar fields with duals defined in terms of the prepotential $\calf$ it follows that this potential itself must transform non-trivially as well. The transformation can be obtained by integration from (\ref{scalartransf}), the result is
\be
\calf^\prime=\calf+\frac{1}{2}\phi_B\left[\transp{C}A\right]^{BC}\phi_C+\phi_B\left[\transp{C}B\right]\ul{B}{C}\phi^C+\frac{1}{2}\phi^B\left[\transp{D}B\right]_{BC}\phi^C\,.\label{pretransf}
\ee
This can be written in the slightly more manageable form as
\be
\calf^\prime-\frac{1}{2}\phi_A^\prime\phi^{\prime A}=\calf-\frac{1}{2}\phi_A\phi^A\,.
\ee

As these formulae make clear, and as is important to note, the action (\ref{bosac}) is in general {\bf not} invariant under duality transformations. It are only the equations of motion following from the action that are invariant. As we will discuss in section \ref{dualandspecial} there exists another variation principle leading to the same equations of motion that is more manifestly duality covariant.

\section{BPS solutions}\label{secBPSsol}
In this section we derive the BPS equations from the Hamiltonian in the standard way and show how they are solved in terms of $2r$ real harmonic functions. The connection between these harmonic functions and the $r$ complex scalars is made through a Legendre transform on the imaginary part of the prepotential, called the Hesse potential.

\subsection{Hamiltonian formalism}
As usual in a Hamiltonian treatment we need to split the coordinates in time and space,  i.e. $x^{\m}=(t,\vec{x})$. Under this split we can then decompose the field strength $F^A_{\m\n}$ into electric and magnetic components $\vec{E}^A$ and $\vec{B}^A$.
The momenta conjugate to $\phi^A$ and $\vec{A}^A$ computed from (\ref{bosac}) are then
\bea
\pi_A&\equiv&\frac{\delta L}{\delta \partial_t\phi^A}=\frac{1}{2}\im\left[\tau_{AB}\right]\partial_t\bar\phi^B\,,\\
\vec{B}_{A}&\equiv&\frac{\delta L}{\delta \partial_t\vec{A}^A}=\im\left[\tau_{AB}(\vec{E}^B+i\vec{B}^B)\right]\,.\label{bd}
\eea
We use the notation $\vec{B}_A$ for the momentum conjugate to $\vec{A}^A$, as one can check that (\ref{bd}) is equal to the magnetic component of $F_A$, as defined in (\ref{doubdef}).

Using these ingredients one computes the Hamiltonian corresponding to (\ref{bosac}):
\bea
H&=&\int d^3x\,\frac{1}{2}\im\left[G_{AB}\right]\left(\vec{E}^A\cdot \vec{E}^B+\vec{B}^A\cdot \vec{B}^B+\partial_t\phi^A\partial_t\bar\phi^B+\grad\phi^A\cdot\grad\bar\phi^B \right)\,.\label{bulkhamiltonian}
\eea
Remember that, as in any U(1) gauge theory, the equations of motion following from this Hamiltonian need to be supplemented by the Bianchi identity and the Gauss constraint:
\be
\grad\cdot \vec{B}^A=0\,,\qquad\qquad\qquad\grad\cdot \vec{B}_A=0\,.\label{gaussbianchi}
\ee
Remark that for the moment we ignore possible sources in these equations as we will focus on the solutions in vacuum regions of space-time. We comment shortly on the introduction of sources in section \ref{comso}.

\subsection{BPS equations}
To find the BPS equations we rewrite the Hamiltonian (\ref{bulkhamiltonian}) as a sum of squares and a total derivative:
\be
H=\int d^3x\,\frac{1}{2}|\vec{E}+i\vec{B}-e^{-i\alpha}\grad\phi|^2+\frac{1}{2}|\partial_t\phi|^2+\grad\cdot\re\left[e^{-i\a}(\phi^A\vec{B}_A-\phi_A\vec{B}^A)\right]\,.\label{squaredhamiltonian}
\ee
Here we used the norm $|X|^2=G_{AB}X^A\cdot\bar X^B$, and introduced an arbitrary constant phase $\alpha$.  The total derivative term gives us the central charge
\be
Z\equiv\int_{\mathrm{S}_\infty^2} d\vec{n}\cdot\left(\phi^A\vec{B}_A-\phi_A\vec{B}^A\right)=\phi^A_\infty q_A-\phi_A^\infty p^A\,.\label{centralcharge}
\ee
In the last step we assumed the scalar fields to asymptote to constant values $\phi_\infty^A$ at large radius and introduced 'electric' and magnetic charges
\be
q_A\equiv \int_{\mathrm{S}_\infty^2} d\vec{n}\cdot\vec{B}_A\,,\qquad\qquad\qquad p^A\equiv \int_{\mathrm{S}_\infty^2} d\vec{n}\cdot\vec{B}^A\,.
\ee 
Due to the positivity of all terms in (\ref{squaredhamiltonian}) and the freedom to choose $\alpha$, the BPS bound follows directly:
\be
H\geq|Z|\,.\label{BPSbound}
\ee
 
Solutions that preserve half of the supersymmetry saturate the bound (\ref{BPSbound}) and are known as BPS solutions. Instead of looking at the Killing spinor equations we can also directly check what the conditions are that the Hamiltonian equals the norm of the central charge. From (\ref{squaredhamiltonian}) it follows that this is the case if and only if the following BPS equations are satisfied:
\bea
\partial_t\phi^A&=&0\,,\label{stat}\\
\arg Z&=&\a\,,\label{phase}\\
\vec{E}^A+i\vec{B}^A-e^{-i\a}\grad\phi^A&=&0\,.\label{BPSeq}
\eea
The first equation tells us that all BPS solutions should be stationary and the second equation fixes the constant $\alpha$ that appears in the last.
This last BPS equation (\ref{BPSeq}) is a set of $r$ first order complex equations. It can be combined with the Gauss and Bianchi equations into $2r$ real Laplace equations, as we will now show.

\subsection{Harmonic conditions}
To start we split (\ref{BPSeq}) in its real and imaginary parts 
\bea
\grad\im\left[e^{-i\a}\phi^A\right]&=&\vec{B}^A\,,\label{BPSB}\\
\grad\re\left[e^{-i\a}\phi^A\right]&=&\vec{E}^A\,.\label{BPSE}
\eea
The first equation (\ref{BPSB}) is readily solved, as together with the the Bianchi identity (\ref{gaussbianchi}) it implies that the imaginary parts of the scalars are harmonic functions:
\be
\Delta\im\left[e^{-i\a}\phi^A\right]=0\,.\label{Lap1}
\ee
The second equation (\ref{BPSE}) can't be treated similarly since in general $\vec{E}^A$ doesn't need to be divergenceless. The way to proceed is to multiply the BPS equation (\ref{BPSeq}) by $\tau_{AB}$:
\be
\tau_{AB}(\vec{E}^B+i\vec{B}^B-e^{-i\a}\grad\phi^B)=\tau_{AB}(\vec{E}^B+i\vec{B}^B)-\grad(e^{-i\a}\phi_A)=0\,.
\ee 
This equation now has as imaginary part
\be
\grad\im\left[e^{-i\a}\phi_A\right]=\vec{B}_A\,.\label{BPSPi}
\ee
Since $\vec{B}_A$ is indeed divergenceless by the Gauss constraint (\ref{gaussbianchi}) we find that also the imaginary parts of the dual scalars are harmonic:
\be
\Delta \im\left[e^{-i\a}\phi_A\right]=0\,. \label{Lap2}
\ee 

So the BPS equations imply $2r$ real Laplace equations (\ref{Lap1}, \ref{Lap2}) on $\mathbb{R}^3$. Given a set of appropriate\footnote{As we discuss in section \ref{comso}, the sources can interact with one another, which will put constraints on what sources are allowed for a BPS solution.} sources these can be solved through standard techniques. The electromagnetic fields are then simply the gradients of these 
harmonic functions. What still has to be clarified however is how the $r$ complex scalars $\phi^A$ are related to the $2r$ real solutions to the Laplace equation. In the next subsection we show that this relation is naturally interpreted as a Legendre transform.

\subsection{Legendre transform}
The motivation for the discussion in this subsection comes form the analysis of the BPS equations made above. It is however important to point out that all formulae in this subsection are valid off-shell. In section \ref{dualandspecial} we will use this to reformulate the theory directly on the level of the action.

The question we want to address is if and how we can express the $r$ complex scalars $\phi^A$ as a function of the $2r$ real scalars\footnote{In principle we can take $\a$ to be an arbitrary phase. If we want to relate this formalism to the BPS solutions we will of course make the identification (\ref{phase})} $\im\left[e^{-i\a}\phi^A\right]$ and $\im\left[e^{-i\a}\phi_A\right]$. To lighten notation and clarify the discussion we will explicitly split up the complex scalars into their real and imaginary parts:
\bea
e^{-i\a}\phi^A&\equiv&X^A+i Y^A\,,\label{defXY}\\
e^{-i\a}\phi_A&\equiv&X_A+i Y_A\,.
\eea
From this split it is clear that the problem to find $\phi^A(Y^A,Y_A)$ reduces to finding $X^A(Y^A,Y_A)$. 

To start remember that $\phi_A$ is defined in terms of $\phi^A$ through the prepotential $\calf$, i.e. $\phi_A\equiv\frac{\partial\calf}{\partial\phi^A}$. Furthermore the prepotential is holomorphic, which implies it satisfies the Cauchy-Riemann conditions
\be
\frac{\partial \calf}{\partial \phi^A}=e^{-i\a}\frac{\partial \calf}{\partial X^A}=-ie^{-i\a}\frac{\partial \calf}{\partial Y^A}\,.
\ee
Combining these two facts one finds the relations between the different real scalars:
\bea
X_A&=&\frac{\partial\im\left[e^{-2i\a}\calf\right]}{\partial Y^A}\,,\\
Y_A&=&\frac{\partial\im\left[e^{-2i\a}\calf\right]}{\partial X^A}\,.\label{rel1}
\eea
The form of these relations shows that we can think of $(Y_A,X^A)$ and $(X_A,Y^A)$ as pairs of conjugate variables and we can trade one component of a pair for the other one by performing a Legendre transform. By our motivation from the BPS equations we will prefer a formulation where the $Y$'s are the $2r$ real variables, so we define the Legendre transform
\be
\cals(Y^A,Y_B)\equiv Y_CX^C-\im\left[e^{-2i\a}\calf(X^A,Y^A)\right]\,.\label{defS}
\ee
In this transform the scalars $X^A$ are now functions of the $2r$ real scalars $(Y^A,Y_A)$, found by inverting (\ref{rel1}). If we consider $\cals$ as a given, then we can find those functions simply by taking a derivative:
\be
X^A=\frac{\partial \cals(Y^B,Y_C)}{\partial Y_A}\,.\label{xups}
\ee
This formula provides, at least formally, the solution to our problem, we can now write
\be
\phi^A(Y^B,Y_C)=e^{i\a}\left(\frac{\partial \cals(Y^B,Y_C)}{\partial Y_A}+i Y^A\right)\,.\label{phiY}
\ee

This expression (\ref{phiY}) is exactly what we need to complete our solution of the BPS equations, since these simply imply that the $Y$'s are all harmonic (\ref{Lap1}, \ref{Lap2}) . So given the appropriate harmonic functions one can plug them into (\ref{phiY}) to find the corresponding complex scalars.

Finally let us point out that one can also directly compute the dual scalars in terms of the $Y$ variables:
\bea
X_A&=&-\frac{\partial \cals(Y^B,Y_C)}{\partial Y^A} \label{xdowns}\qquad
\phi_A=e^{i\a}\left(-\frac{\partial \cals(Y^A,Y_A)}{\partial Y^A}+i Y_A\right)\,.
\eea
The minus sign that appears here is important and is directly related to the symplectic structure on moduli space, as we will discuss in the next section.

\section{The Hesse potential: duality and special geometry}\label{dualandspecial}
In our study of the BPS equations in the previous section we were naturally led to introduce a function $\cals$ as the Legendre transform of the imaginary part of the prepotential $\calf$, see (\ref{defS}). In this section we will show that the Hesse potential $\cals$ is invariant under duality and that the special geometry of the theory can be elegantly expressed in terms of it. Most of this will be review of the discussions in \cite{Freed:1997dp, Hitchin:2005uu, Ferrara:2006js}. In the second part of this section we show how the theory can be expressed in a duality covariant description that seems most naturally connected to the BPS equations and solutions and we comment on the constraints on sources.

\subsection{The Hesse potential}
As we will show in this subsection, $\cals$ plays a role in special geometry that is very analogous to that of the prepotential $\calf$, the difference being that $\calf$ makes the complex structure of the manifold manifest while a description in terms of $\cals$ makes the symplectic and duality structure manifest. This will present itself in two ways, first of all we will show that $\cals$ is duality invariant and secondly that in Darboux coordinates the metric is simply given by the Hessian of $\cals$. Due to this last property $\cals$ has been called the {\it Hesse potential} \cite{Cardoso:2006bg, Cardoso:2010gc}. 

To check the duality invariance of $\cals$ only takes a few lines of algebra. It follows from combining the definition (\ref{defS}) in terms of the prepotential $\calf$, with the transformation properties (\ref{scalartransf}, \ref{pretransf}):
\bea
\cals^\prime&=&Y_A^\prime X^{\prime A}-\im\left[e^{-2i\a}\calf^\prime\right]\\
&=&\cals+\frac{1}{2}\left(X_AY^A-X^AY_A\right)-\frac{1}{2}\left(X_A^\prime Y^{\prime A}-X^{\prime A}Y_A^\prime\right)\\
&=&\cals\,.
\eea
The last step follows because $X_AY^A-X^AY_A=X_A^\prime Y^{\prime A}-X^{\prime A} Y_A^\prime$ is an invariant under the symplectic duality group Sp($2r,\mathbb{R}$). This is most clear if we introduce doublet notation and a symplectic form:
\be
X^a=\begin{pmatrix}X_A\\X^A\end{pmatrix}\,,\qquad Y^a=\begin{pmatrix}Y_A\\Y^A\end{pmatrix}\,,\qquad\quad\Omega=\begin{pmatrix}0&-1\\1&0\end{pmatrix}\,.
\ee
In this notation $X_AY^A-X^AY_A=\Omega_{ab}Y^aX^b$ which is manifestly invariant under symplectic transformations. Note that this invariant is nothing but the K\"ahler potential (\ref{Kahlerpot}), i.e. $\calk=\Omega_{ab}X^aY^b$.

The invariance of $\cals$ under duality rotations suggests that it is closely related to the symplectic structure of the theory. Indeed, as we will now describe it plays a role similar to the prepotential, in that its second derivatives give the metric on the scalar manifold. However, since $\cals$ is a function of the $2r$ real coordinates $Y^a$ instead of the $r$ complex $\phi^A$, the matrix of second derivatives is nothing but the Hessian matrix of $\cals$.

To make this explicit one can start from the scalar metric in complex coordinates (\ref{scmetr}) and rewrite it using the relations (\ref{xups}, \ref{xdowns}):
\bea
G_{AB}d\phi^Ad\bar\phi^B&=&\im\left[d\phi_Ad\bar\phi^A\right]\nonumber\\
&=&dY_AdX^A-dY^AdX_A\label{scredef}\\
&=&\frac{\partial^2\cals}{\partial Y^a\partial Y^b}dY^adY^b\,.\nonumber
\eea
In summary the scalar metric in the $2r$ real coordinates $Y^a$ is indeed the Hessian matrix of $\cals$, we will use the following notation:
\be
\calg_{ab}\equiv \frac{\partial^2\cals}{\partial Y^a\partial Y^b}\,.\label{hesmet}
\ee
What is interesting is that the metric $\calg_{ab}$ is actually a rather familiar object. Remember that we have both a symplectic form $\Omega_{ab}$ and a natural complex structure $\cali^{a}{}_b$ defined in (\ref{self-dual}), that are furthermore compatible, i.e. $\Omega_{ab}=\Omega_{cb}\cali^c{}_{a}\cali^{d}{}_b$. This implies that together they define a metric: 
\be
\calg_{ab} =\Omega_{ac}\cali^{c}{}_b\quad\Rightarrow\quad \calg=\begin{pmatrix}G^{-1}&\ & -G^{-1}\Theta \\
 -\Theta G^{-1} &\ &G+\Theta G^{-1}\Theta\end{pmatrix}\,.\label{dualtmetric}
\ee
As we already implied in our notation, this metric is nothing but the Hessian metric (\ref{hesmet}). This was shown in e.g. \cite{Cortes:2001qd, Cortes:2005uq, Ferrara:2006js}, but for completeness and since we will use this result later on in the paper, we have added a derivation of this fact using our notation in appendix \ref{appmetr}.

\subsection{Duality covariant formalism}
In the discussion of the BPS equations and their solutions in section \ref{secBPSsol} we found that those are most naturally expressed in terms of the $2r$ real scalars $Y^a$ and the magnetic fields $\vec{B}^a$. To realize this we however had to first perform some algebra and derive the function $\cals$ from the prepotential $\calf$. Using the results of the previous subsection we will now present an alternative formulation of the theory where these variables appear as the fundamental degrees of freedom in the action and the simplest form of the BPS equations follows immediately. By construction this new action will be manifestly covariant under duality. In appendix \ref{appdual} we show in detail that the new formulation leads to the same equations of motion\footnote{Although in this paper we are mainly interested in BPS solutions and equations, the equivalence of the theories is checked for the complete set of second order equations of motion without any further assumptions on BPS or other conditions.} as the more standard description in terms of the action (\ref{bosac}).

In the original description (\ref{bosac}) there are $r$ complex scalars $\phi^A$ and $r$ 4d gaugefields $A_\mu^A$ as fundamental variables. The gauge potentials only enter the action through the field strengths $F^A=dA^A$, which is thus manifestly invariant under U(1)$^r$ gauge transformations $A^A_\mu\rightarrow A^A_\m+\partial_\m\Lambda$. In this formulation duality transformations act in a nontrivial non-local way. For such Abelian gauge theories there exists a nice alternative formulation that makes duality manifestly covariant and furthermore implements it as a local transformation \cite{Deser:1976iy, Schwarz:1993vs}. Combined with the field redefinition  (\ref{scredef}) in the scalar sector this formalism is naturally extended to any $\caln=2$ low energy gauge theory.

In this alternative description we will take as fundamental variables $2r$ real scalars $Y^a$, $2r$ 3d vectors $\vec{A}^a$ and $2r$ functions $A_0^a$. Note the naive 'doubling of degrees of freedom' in the electromagnetic sector with respect to the usual formulation (\ref{bosac}). Using the vector potentials we define the following magnetic and electric fields:
\be
\vec{B}^a\equiv\rot \vec{A}^a\,,\qquad \vec{E}^a\equiv\partial_t \vec{A}^a-\grad A_0^a\,.
\ee
By construction these satisfy the following 'Bianchi-identities':
\be
\grad\cdot \vec{B}^a=0\,,\qquad \partial_t\vec{B}^a=\rot\vec{E}^a\,. \label{dualbianch}
\ee
Given the symplectic form $\Omega_{ab}=-\epsilon_{ab}$ and a scalar dependent metric $\calg_{ab}(Y)$, we write down the following Lagrangian:
\be
L=-\frac{1}{2}\left(\Omega_{ab}\vec{E}^a\cdot\vec{B}^b+\calg_{ab}\vec{B}^a\cdot\vec{B}^b+\calg_{ab}\grad Y^a\cdot\grad Y^b-\calg_{ab}\partial_tY^a\partial_tY^b\right)\,.\label{dualcovlag}
\ee
First of all it is interesting to note that this Lagrangian has the following two gauge invariances
\bea
A_0^a\rightarrow A_0^a+\Lambda_0^a&\quad \Rightarrow\quad& \vec{E}^a\rightarrow \vec{E}^a-\grad\Lambda_0^a\,,\ \quad L\rightarrow L+\frac{1}{2}\grad\cdot(\Omega_{ab}\Lambda_0^{a}\vec{B}^b)\,,\label{gaugtransfo0}\\
\vec{A}^a\rightarrow\vec{A}^a+\grad \Lambda^a&\quad\Rightarrow\quad&\vec{E}^a\rightarrow \vec{E}^a+\partial_t\grad\Lambda^a\,,\ \quad L\rightarrow L-\frac{1}{2}\grad\cdot(\Omega_{ab}\partial_t\Lambda^{a}\vec{B}^b)\,.
\eea
Remark that the electric fields in this formulation are {\bf not} gauge invariant, contrary to what we are used to in Maxwell's formulation of electromagnetism. Furthermore the first line shows us that the fields $A_0^a$ are pure gauge and can be put to an arbitrary value. This is consistent with the fact that they appear as a total derivative in the Lagrangian (\ref{dualcovlag}), hence varying with respect to them will not lead to any non-trivial field equation. Again this is different from the standard Maxwell action.

Furthermore under the Sp$(2r,\mathbb{R})$ duality group all fields now simply transform in the vector representation, i.e. for $M^{a}{}_{b}\in$ Sp$(2r,\mathbb{R})$ we have $f^a\rightarrow M^{a}{}_{b}f^{b}$. Note that this is also true for the vector potentials $\vec{A}^a$, so these now transform in a manifestly local way. The Lagrangian (\ref{dualcovlag}) is in general not invariant under these duality transformations, only for special choices of $\calg_{ab}$ this will be the case. However, it immediately follows from the covariance of all fields in the action that the set of equations of motion is invariant under duality transformations.

We deliberately used the same notation for the fields appearing in the Lagrangian (\ref{dualcovlag}) as for those we introduced earlier in the text. As is shown in appendix \ref{appdual}, under these identifications the equations of motion following from (\ref{dualcovlag}) are equivalent to those of the action (\ref{bosac}), and so both Lagrangians are nothing but different descriptions of the same physical theory.

One can derive the BPS equations directly in this new description. To do so let us again go to the Hamiltonian formulation, the momenta conjugate to the new variables are 
\be
\vec{\Pi}_a=\frac{\delta L}{\delta \partial_t \vec{A}^a}=-\frac{1}{2}\Omega_{ab}\vec{B}^b\,,\qquad \pi_a=\calg_{ab}\partial_tY^b\,.
\ee
The Hamiltonian density is then computed to be
\be
\calh=\frac{1}{2}\calg_{ab}\left(\vec{B}^a\cdot\vec{B}^b+\grad Y^a\cdot\grad Y^b+\partial_tY^a\partial_tY^b\right)\,.\label{hardens2}
\ee
Note that contrary to the Lagrangians the two Hamiltonians (\ref{bulkhamiltonian}, \ref{hardens2}) are actually equal on-shell, as expected from their physical interpretation as the energy density of the solution.
We can also rewrite the Hamiltonian density (\ref{hardens2}) as a sum of squares plus a total derivative
\be
\calh=\frac{1}{2}\calg_{ab}(B^a-\grad Y^a)(B^b-\grad Y^b)+\frac{1}{2}\calg_{ab} \partial_t Y^a\partial_t Y^b+\grad\cdot\left(\partial_a\cals\,\vec{B}^a\right)\,.\label{ham2totsq}
\ee
This expression immediately implies the following bound:
\be
H\geq\int_{S^2_\infty}\partial_a\cals\,\vec{B}^a\cdot d\vec{n}\,.
\ee
In the next section we will discuss how this bound is nothing but the standard BPS bound in our new notation. Demanding the bound to be saturated gives the BPS equations, that now take a very simple form:
\be
\partial_tY^a=0\,,\qquad\vec{B}^a=\grad Y^a\,,\qquad \Delta Y^a=0\,.\label{dualcovBPSeq}
\ee
Here the third equation follows from the second by the Bianchi identities (\ref{dualbianch}).
Of course the equations (\ref{dualcovBPSeq}) are equivalent to the BPS equations derived in section \ref{secBPSsol}.

\subsection{A comment on sources}\label{comso}
Since the BPS solutions are simply determined in terms of harmonic functions one might naively assume that one can take linear superpositions of any number of solutions. This is however {\bf not} the case when these carry mutually non-local charges. The subtlety comes from the precise treatment of the sources to these harmonic functions. We leave a general and detailed discussion to future work, but illustrate the issue in the case of static pointlike sources. In this case we need to modify the equations (\ref{dualcovBPSeq}) to $\Delta Y^a=\sum_i \Gamma^a_i\delta^3(\vec{x}-\vec{x_i})$, which gives
\be
Y^a=\sum_i \frac{\Gamma^a_i}{|\vec{x}-\vec{x}_i|}+Y^a_\infty\,.
\ee
To fully take this effect into account we also need to add the energetic contribution of the sources to the Hamiltonian (\ref{hardens2}), i.e.
\be
\calh_{\mathrm{tot}}=\calh+\sum_i |\calz_i|\delta^3(\vec{x}-\vec{x_i})\,,\qquad \calz_i=\Omega_{ab}\phi^a(\vec{x})\Gamma^b_i\,. \label{sourceadd}
\ee
When we rewrite $\calh$ as a total square as in (\ref{ham2totsq}), the use of the Bianchi identities now give rises to an additional term $-\sum_i \re\left[e^{i\a}\calz_i\right]\,\delta^3(\vec{x}-\vec{x_i})$\,. To have a solution that saturates the BPS bound this term needs to cancel against the source contribution in (\ref{sourceadd}), which leads to constraints on the allowed relative positions of the sources:
\be
\sum_j \frac{\Omega_{ab}\Gamma_i^a\Gamma_j^b}{|\vec{x}_i-\vec{x}_j|}=\Omega_{ab}Y^a_\infty\Gamma^b_i\,.
\ee
These are the well known stability conditions \cite{Denef:2000nb} that lead to the non-trivial bound state structure of generic $\caln=2$ BPS solutions. Note that apart from this algebraic constraint on possible source locations our analysis before was completely general and is valid everywhere in the vacuum away from the sources. In the rest of the paper we focus again on this vacuum behavior of the solutions.

\section{Generalized attractor equations and Hesse flow}\label{genflow}
BPS solutions to low energy $\caln=2$ supergravity and gauge theories have been especially well studied in the spherically symmetric case. In this case there is only dependence on a single radial coordinate, and it turns out that the BPS equations imply a simple radial flow for a local version of the central charge. This phenomenon is probably best known in the supergravity context, where it goes under the name of the attractor mechanism. This result gives a nice physical ‌interpretation to the BPS solution. We start out with boundary conditions that set the central charge at infinity, and as we then move radially inward this central charge becomes smaller and smaller, reaching its minimum at a certain radius $r_\star$.

In this subsection we want to show how this formalism generalizes to the cases without spherical symmetry, where in general there is dependence on all three spatial coordinates. As we will argue, the standard notion of local central charge will need to be generalized to a central charge vector field. We we will then show that the BPS equations imply that the real part of this central charge vector field becomes the gradient of a potential, and that this potential is none other than the Hesse potential $\cals$ evaluated on the solution. Moreover, we will see that this equation is only one in a set of four, that take the form of Maxwell's equations. The role of electric charge will be played by the energy density in the solution, while there is also a magnetic component to the central charge vector field that couples to the momentum current in the solution.

\subsection{The central charge vector field}
So far we have defined the central charge of a solution (\ref{centralcharge}) as the complex number 
\be 
Z=p^A\phi_A^\infty-q_A\phi^A_\infty\,.
\ee As is well known \cite{Witten:1978mh}, this is the quantity showing up in the superalgebra of the theory. More generally we can define a bilinear central charge function that maps any duality covariant real vector $V^a$ to a complex number for each value of the scalar moduli $\phi^A$, i.e. 
\be
z: \mathbb{R}^{2r}\times \calm \rightarrow \mathbb{C} : (V,\phi)\mapsto z(V,\phi)\equiv\Omega_{ab}\phi^aV^b\,,\label{ccfunction}
\ee
where we again used the notation $\phi^a=(\phi^A,\partial_A\calf)$ as in (\ref{scalartransf}). So in our terminology the {\it central charge} of a solution is nothing but the image under the {\it central charge function}, of the total charge $\Gamma^a=(q_A,p^A)$ and the asymptotic values of the scalars in that solution, i.e. $Z=z(\Gamma,\phi_\infty)$. In previous studies of BPS solutions it became clear that a closely related notion, that of a {\it local central charge}, is useful in understanding the physics of the solution. Given a scalar field configuration, i.e. $\phi^A(\vec x)$ and a total charge $\Gamma^a$, one defines
\be
\calz(\vec{x})\equiv z(\Gamma,\phi(\vec{x}))=\Omega_{ab}\phi^a(\vec{x})\Gamma^b\,.
\ee
By definition the local central charge evaluated at infinity equals the global central charge, i.e. $\lim_{\lambda\rightarrow\infty}\calz(\lambda\vec{x})=Z$. In the spherically symmetric case the local central charge only depends on the radial coordinate, i.e. $\calz=\calz(r)$. It is then a well known result that the BPS equations have a simple and very physical interpretation in terms of this central charge. In the gauge theory setting the BPS equations imply \cite{Chalmers:1996ya, Denef:2000nb}:
\be
\im\left[e^{i\a}\calz\right]=0\qquad\mbox{and}\qquad \partial_r |\calz|=\frac{1}{2}G^{AB}\partial_A\calz\bar\partial_B\bar\calz\geq0\,.\label{attreq}
\ee
Remember that $\a=\arg Z$, so the first equation implies the local central charge has a position independent constant phase. The second equation then tells us that the norm of the central charge decreases with respect to the radial coordinate. A more precise analysis reveals that the minimum is reached at finite radius. What is important is that these equations only hold in the spherically symmetric case. As far as we are aware there exist no simple to interpret equations for the local central charge when there is no spherical symmetry and non-mutually local charges are present. 

We will now propose a more general notion of 'local central charge', for which a nice physical interpretation of the BPS equations will appear for all BPS solutions and that in the spherically symmetric case reduces to (\ref{attreq}). 

When looking to generalize $\calz=z(\Gamma,\phi(\vec{x}))$ the first thing that comes to mind is to also replace the first argument by a space-time dependent quantity. A simple guess would be to replace $\Gamma^a\rightarrow Y^a(\vec{x})$, this is interesting as this new 'local central charge', $\tilde\calz(\vec{x})\equiv z(Y(\vec{x}),\phi(\vec{x}))=\Omega_{ab}\phi^a(\vec{x})Y^b(\vec{x})$ has a fixed phase by construction:
\be
\im\left[e^{-i\a}\tilde \calz\right]=\Omega_{ab}Y^aY^b=0\,. \label{cstph}
\ee 
So $\tilde\calz$ seems to at least generalize the first equation of (\ref{attreq}) to generic solutions. However there seems to be no nice equation describing the spatial dependence of the norm of $\tilde\calz$, which would generalize the second equation in (\ref{attreq}). Furthermore there is the additional problem that in the special case with spherically symmetry it becomes clear that $\tilde\calz$ is a physically different object than $\calz$. Spherical symmetry implies that $Y^a=\frac{\Gamma^a}{r}+Y^a_\infty$ and so $\tilde\calz=\frac{\calz}{r}+z(Y_\infty,\phi(\vec{x}))$ in that case. The second term is non-trivial and so $\tilde\calz$ satisfies different equations than $\calz$, even in the simple spherically symmetric case. Hence this is not the generalization we are looking for.

A more fruitful proposal is to replace $\calz$ by a 3-vector field $\vec{\calz}$ that we define as follows:
\be
\vec{\calz}(\vec{x})\equiv z(\vec{B}(\vec{x}),\phi(\vec{x}))=\Omega_{ab}\phi^a(\vec{x})\vec{B}^b(\vec{x})\,.\label{vecz}
\ee
Note that here $\vec{B}^a$ is the symplectic vector containing both the electric and magnetic fields (\ref{doubdef}). A first good property of this {\it central charge vector field}\footnote{Note that this vectorfield is analogous to the graviphoton field strength in supergravity, as it is essentially the unique duality invariant contraction of the scalars and electro-magnetic fields. We thank T. Mohaupt for pointing this out to us.} $\vec{\calz}$, that justifies to call it a generalization of $\calz$, is that in the spherically symmetric case both are essentially the same. Assuming spherically symmetry the electromagnetic fields have the form $\vec{B}^a=\frac{\Gamma^a}{r^3}\vec{r}$ and thus $\vec{\calz}=\frac{\calz}{r^3}\vec{r}$. So in this special case $\vec\calz$ shares the same constant phase with $\calz$ and the norms of the two quantities are proportional by a monotonous radial factor and thus exhibit the same attractor behavior. 

In the more general non-spherically symmetric case, especially when multiple non-mutually local charges are present, $\calz$ is no longer simply related to $\vec\calz$. However, as we will show explicitly in the next subsection, it is $\vec\calz$ that satisfies simple first order equations in any context, with no additional symmetry assumptions. This suggests $\vec{\calz}$ is really the natural local generalization of the global central charge $Z$ defined at infinity.

\subsection{Hesse flow and central magnetism}
Let us now show how one can generalize the equations (\ref{attreq}) to cases without spherical symmetry and non-mutually-local charges, making use of the central charge vector field $\vec{\calz}$ defined in (\ref{vecz}).
Even in this more general case, if we assume all charges to be contained in some finite region, then near infinity the field strengths will asymptotically approach those of a spherically symmetric distribution. So it follows that the phase of the central charge vector field at infinity is equal to that of the global central charge (\ref{centralcharge}). In formulas 
\be
\lim_{\l\rightarrow\infty}\im \left[e^{-i\a}\vec{\calz}(\l \vec x)\right]=0\qquad\mbox{with}\quad \a=\arg(Z)\,.
\ee
But once we move away from the boundary at infinity 'the' phase of $\vec{\calz}$ will become position dependent. To be more precise, each of the three components actually develops an independent position dependent phase. So the description in terms of the norm and the phase, that we used to formulate (\ref{attreq}), seems not so useful for generic BPS solutions. Rather it will prove natural to split up the complex central charge vector $\vec{\calz}$ in its real and imaginary components, $\re [e^{-i\a}\vec{\calz}]$ and $\im [e^{-i\a}\vec{\calz}]$. Note that this split becomes related to the norm and phase when the solution has spherical symmetry, then one finds $\re [e^{-i\a}\vec{\calz}]=\frac{|\calz|}{r^3}\vec{r}$ and $\im [e^{-i\a}\vec{\calz}]=0$.

Let us start our analysis with the real part. Simply writing out its definition from (\ref{vecz}) in terms of the symplectic coordinates $Y^a$ and their Hesse potential $\cals$ that we introduced in the previous sections, one finds $\re [e^{-i\a}\vec{\calz}]=\vec{B}^a\partial_a \cals$. In this form it becomes a simple observation that the BPS equations $\vec{B}^a=\vec{\nabla}Y^a$ imply a gradient flow equation:
\be
\rez=\vec{\nabla}\cals\,.\label{Hessflow}
\ee  
Note that here the Hesse potential $\cals$ is to be considered as a function on $\mathbb{R}^3$, evaluated as $\cals(Y(\vec{x}))$. Equation (\ref{Hessflow}) shows that the Hesse potential, apart from being an ingredient in the special geometry of the theory, also gets a physical interpretation on BPS solutions, as a potential for the real part of the central charge vector field. This gradient flow in $\cals$, which we will refer to as Hesse flow, becomes the well know attractor flow in the spherically symmetric case, but is more general and as we just showed holds for all BPS solutions irrespective of symmetries or charges. 

The Hesse flow equation (\ref{Hessflow}) is however only the first manifestation of a bigger underlying structure characterizing BPS solutions. A rather trivial consequence is that for BPS solutions the real part of the central charge vector field is irrotational, i.e. 
\be
\vec{\nabla}\times\rez=0\,.
\ee
More interesting is to compute its divergence. Evaluating it for vacuum BPS solutions one finds
\be
\vec{\nabla}\cdot\rez=\calh\,.
\ee 
Here $\calh$ is the energy density (\ref{hardens2}) of the BPS solution. Note that this equation is not at all unexpected, it is simply the local version of the well known BPS condition $H=|Z|$, and it is an immediate consequence of the expression (\ref{ham2totsq}). Combining it with the Hesse flow equation (\ref{Hessflow}) reveals an intriguing relation however:
\be
\Delta\cals=\calh\,.
\ee
Here we see that the Hesse potential evaluated on a BPS solution satisfies Poisson's equation sourced by the energy density. In physical terms, for a BPS solution the Hesse potential is non other than the gravitational potential! Furthermore (\ref{Hessflow}) then gives the real part of the central charge vector field the interpretation of the gravitational field. 

After uncovering these rather elegant equations for the real part of the central charge vector field, let us analyse the imaginary part. Although this is trivially zero in spherically symmetric solutions, that is not the case in general. Again one starts from the definition (\ref{vecz}) and rewrites things in terms of the real scalars $Y^a$: $\im[e^{-i\a}\vec{\calz}]=\Omega_{ab}Y^a\vec{B}^b$. Since the BPS equations read $\vec{B}^a=\grad Y^a$ and the $Y^a$ are harmonic it follows immediately that
\be
\vec{\nabla}\cdot\imz=0\,.
\ee
Furthermore the BPS equations imply an elegant expression for the curl as well. One computes
\be
\vec{\nabla}\times\imz=2\vec{\calp}\,.
\ee
Here $\vec{\calp}=\frac{1}{2}\Omega_{ab}\vec{B}^a\times\vec{B}^a$ is nothing but the Poynting vector field of the solution.

A very nice structure unfolds when we put all these expressions together. The calculations above show that the BPS equations imply that the central charge vector field (\ref{vecz}) satisfies the following equations:
\bea
\vec{\nabla}\cdot\re \left[\frac{e^{-i\a}}{2}\vec{\calz}\right]=\calh_{\mathrm{e.m.}}\,,&\qquad&\vec{\nabla}\cdot\im \left[\frac{e^{-i\a}}{2}\vec{\calz}\right]=0\,,\label{deMax1}\\
\vec{\nabla}\times\re \left[\frac{e^{-i\a}}{2}\vec{\calz}\right]=0\,,&\qquad&\vec{\nabla}\times\im \left[\frac{e^{-i\a}}{2}\vec{\calz}\right]=\vec{\calp}\,.\label{deMax2}
\eea
These are none other than Maxwell's equations, with (half of) the real part of the central charge vector field playing the role of 'electric field', while the imaginary part behaves magnetically. Note that the role of 'electric charge density' is played by the energy density of the electromagnetic fields in the theory, $\calh_{e.m.}=\frac{1}{2}\calh$ for BPS solutions. The current $\vec{\calp}$ that sources the 'magnetic' imaginary part is none other than the momentum current. Indeed these equations are Lorentz invariant as $\calh_{e.m.}$ and $\vec{\calp}$ nicely combine in the energy-momentum 4-vector.

It is straightforward to check that in the spherically symmetric case the magnetic equation becomes trivial, while the electric flow equation becomes the attractor equation (\ref{attreq}) (up to a trivial radial rescaling). 

\section*{Acknowledgements}
It is a pleasure to thank F. Denef, T. Mohaupt and M. Shigemori for interesting discussions. This work is partially supported by TUBITAK Grant 107T896. During part of this work the author was affiliated to the NHETC at Rutgers University and supported by the DOE under grant DE-FG02-96ER40959. He also enjoyed the hospitality of the Aspen Center for Physics where he was supported in part by the National Science Foundation under Grant No. 1066293.

\appendix
\section{Duality covariant metric and the Hesse potential}\label{appmetr}
In this appendix we show that the duality covariant metric introduced in (\ref{dualtmetric}) is the Hessian of the potential $\cals$ defined in (\ref{defS}), i.e. 
$\calg_{ab}=\frac{\partial^2\cals}{\partial Y^a \partial Y^b}$, with $\cals=Y_AX^A-\im\left[e^{-2i\a}\calf\right]$ and 
\be
\calg=\begin{pmatrix}G^{-1}&\ & -G^{-1}\Theta \\
 -\Theta G^{-1} &\ &G+\Theta G^{-1}\Theta\end{pmatrix}\,.\label{dualtmetricapp}
\ee
Before we start the derivation of this fact, let us for convenience first summarize some definitions and identities from the main body of the paper. Remember that 
$X^A=\frac{\partial \cals}{\partial Y^A}$, $X_A=-\frac{\partial \cals}{\partial Y^A}$
and we use the notation $\tau_{AB}=\frac{\partial\calf}{\partial\phi^A\partial\phi^B}=\Theta_{AB}+i G_{AB}$. Furthermore a useful form of the Cauchy-Riemann equations for any holomorphic function $H(\phi)$, $\phi=e^{i\a}(X+iY)$ is
\be
\frac{\partial H}{\partial \phi^A}=e^{-i\a}\frac{\partial H}{\partial X^A}=-ie^{-i\a}\frac{\partial H}{\partial Y^A}\,.
\ee
Combining the different identities above one finds that
\bea
\frac{\partial Y_A}{\partial Y^B}&=&\frac{\partial X_A}{\partial X^B}=\Theta_{AB}\,,\\
\frac{\partial Y_A}{\partial X^B}&=&\frac{\partial X_A}{\partial Y^B}=G_{AB}\,.
\eea
Let us furthermore consider the coordinate transformation $(X^A,Y^A)\rightarrow (Y_A,Y^A)$, the Jacobian of this transformation and it's inverse are given by:
\be
J=\begin{pmatrix}\frac{\partial Y_A}{\partial X^B}&\frac{\partial Y_A}{\partial Y^B}\\
0&\delta^A_B\end{pmatrix}\,,\qquad\qquad J^{-1}=\begin{pmatrix}\frac{\partial X^A}{\partial Y_B}&\frac{\partial X^A}{\partial Y^B}\\
0&\delta^A_B\end{pmatrix}\,.
\ee
Because of the fact that $JJ^{-1}=1$ the following relations follow:
\bea
\frac{\partial Y_A}{\partial X^B}\frac{\partial X^B}{\partial Y_C}&=&\delta^C_A\,,\\
\frac{\partial Y_A}{\partial X^B}\frac{\partial X^B}{\partial Y^C}&=&-\frac{\partial Y_A}{\partial Y^C}\,.
\eea

We can now use the results above to compute the Hessian of $\cals$ in terms of the original couplings $\tau$. It is a matter of simple algebra to check that
\be
\frac{\partial^2 \cals}{\partial Y_A\partial Y_B}=G^{AB}\,,\qquad\qquad \frac{\partial^2 \cals}{\partial Y_A\partial Y^B}=-\left[G^{-1}\Theta\right]^{A}{}_{B}\,.
\ee
Computing the remaining components of the Hessian is a little more subtle, one needs to take into account that
\be
\frac{\partial^2 \cals}{\partial Y^A\partial Y^B}=\frac{d X_A}{d Y^B}=\frac{\partial X_A}{\partial Y^B}+\frac{\partial X_A}{\partial Y_C}\frac{\partial Y_C}{\partial Y^B}\,.
\ee
We can now again apply the various identities above to find:
\be
\frac{\partial^2 \cals}{\partial Y^A\partial Y^B}=G_{AB}+\left[\Theta G^{-1}\Theta\right]_{AB}\,.
\ee
Summarized in matrix notation the results look as follows
\be
\mathrm{Hess}(\cals)=\begin{pmatrix}\frac{\partial^2 \cals}{\partial Y_A\partial Y_B}&\frac{\partial^2 \cals}{\partial Y_A\partial Y^B}\\\frac{\partial^2 \cals}{\partial Y^A\partial Y_B}&\frac{\partial^2 \cals}{\partial Y^A\partial Y^B}\end{pmatrix}=\begin{pmatrix}G^{AB}&\quad& -G^{AC}\Theta_{CB} \\
 -\Theta_{AC} G^{CB} &\quad& G_{AB}+\Theta_{AC} G^{CD}\Theta_{DB}\end{pmatrix}\,.
\ee
By comparing with (\ref{dualtmetricapp}) we see that indeed, as we set out to show, Hess$(\cals)=\calg$. Using the index notation $Y^a=(Y_A,Y^A)$ this can be written as $\calg_{ab}=\frac{\partial^2 \cals}{\partial Y^a\partial Y^b}$.

\section{Equivalence of duality covariant description}\label{appdual}
In the main text we introduced two Lagrangians, (\ref{bosac}) and (\ref{dualcovlag}), (and their corresponding Hamiltonians, (\ref{bulkhamiltonian}) and (\ref{hardens2})) to describe a single low energy $\caln=2$ gauge theory. The Lagrangians are functions of different variables and are extremized with respect to two different variational principles. But as we will explicitly show in this appendix, they lead to equivalent equations of motion if one identifies their fundamental fields appropriately. For the electromagnetic sector of our theories both the formulation of the duality covariant Lagrangian and the check on its equations of motion are a rather straightforward generalization of the formalism of \cite{Deser:1976iy, Schwarz:1993vs}. In our particular case there is however also a scalar sector that couples non-trivially to the electromagnetic fields and we show that even including those, the complete sets of equations of motion of the two Lagrangians are equivalent.

\subsection{The two descriptions}
Let us for clearness start by comparing the two Lagrangians and their fundamental variables, at the same time splitting them up into pieces that play different roles. 

The first, most standard and manifestly Lorentz invariant, Lagrangian (\ref{bosac}) is
\bea
L_1(\phi^A,A^A_{\mu})&=&L_1^{\mathrm{e.m.}}(\phi^A,A^A_{\mu})+L_1^{\mathrm{sc. kin.}}(\phi^A)\,,\label{lag1}\\
L_1^{\mathrm{e.m.}}&=&-\frac{1}{4}G_{AB}F^A_{\m\n}F^{B\,\m\n}-\frac{1}{4}\Theta_{AB}F^A_{\m\n}\star\!F^{B\,\m\n}\,,\\
L_1^{\mathrm{sc. kin.}}&=&-\frac{1}{2}G_{AB}\partial_\m\phi^A\partial^\m\bar\phi^B\,.
\eea
Its fundamental variables are $r$ complex scalars $\phi^A$ and $r$ 4d gaugefields $A^A_\mu$. We will call this theory 'description 1'.

The second, manifestly duality covariant, Lagrangian (\ref{dualcovlag}) is
\bea
L_2(Y^a,\vec{A}^a,A_0^a)&=&L_2^{\mathrm{e.m.}}(\vec{A}^a,A_0^a)+L_2^{\mathrm{m.}+\mathrm{sc.}}(\vec{A}^a,Y^a)+L_2^{\mathrm{sc. kin.}}(Y^a)\,,\label{lag2}\\
L_2^{\mathrm{e.m.}}&=&-\frac{1}{2}\Omega_{ab}\vec{E}^a\cdot\vec{B}^b\,,\\
L_2^{\mathrm{m.}+\mathrm{sc.}}&=&-\frac{1}{2}\calg_{ab}\vec{B}^a\cdot\vec{B}^b\,,\\
L_2^{\mathrm{sc. kin.}}&=&\frac{1}{2}\calg_{ab}\partial_tY^a\partial_tY^b-\frac{1}{2}\calg_{ab}\grad Y^a\cdot\grad Y^b\,.
\eea
Here the fundamental variables are $2r$ real scalars $Y^a$, $2r$ 3d vector potentials $\vec{A}^a$ and $2r$ potentials $A_0^a$. Note that these last are pure gauge and their variation doesn't lead to a non-trivial field equation, see section \ref{dualandspecial} for some details. This second theory we will refer to as 'description 2'.

We will now show that the field equations of these two theories are exactly the same and the two are different descriptions of the same physics, provided we make the following identifications:
\bea
Y^a&=&\begin{pmatrix}\im\left[e^{-i\a}\phi_A\right]\\ \im\left[e^{-i\a}\phi^A\right]\end{pmatrix}\,,\nonumber\\
\vec{B}^a&=&\begin{pmatrix}\vec{B}_A\\ \vec{B}^A\end{pmatrix}\quad\mathrm{with}\quad\vec{B}_A\equiv G_{AB}\vec{E}^B+\Theta_{AB}\vec{B}^B\,,\label{id12}
\\ \calg_{ab}&=&\begin{pmatrix}\left[G^{-1}\right]^{AB}&\ & -\left[G^{-1}\Theta\right]^{A}{}_{B} \\
 -\left[\Theta G^{-1}\right]_{A}{}^B &\ &G_{AB}+\left[\Theta G^{-1}\Theta\right]_{AB}\end{pmatrix}\,.\nonumber
\eea 
It is important to note that for each field and coupling in description 1, we have now identified a corresponding physical field or coupling in description 2. There remains however a field in description 2, i.e. $\vec{E}^a$, that so far (i.e. off-shell) has no counterpart in description 1. We cannot link it to a physical field in description 1 without introducing a constraint, since above we have already associated a field of description 2 to each field of description 1. As we will discuss, the equations of motion following from the respective Lagrangians provide exactly this constraint and so on-shell the two theories are equivalent. Furthermore, as we will stress and explain below, it is important that the identifications (\ref{id12}) are made only after the respective Lagrangians have been varied.

\subsection{Electromagnetic field equations}
Let us start by showing that the field equations for the electromagnetic fields are equivalent.

In description 1 the independent fundamental electromagnetic fields are $\vec{E}^A$ and $\vec{B}^A$. As we described in the main text in section \ref{subsecdualstr}, the field equations of description 1 take the following form:
\begin{itemize}
\item {\bf Bianchi identities}
\be
dF^A=0\quad\Leftrightarrow\quad \grad\cdot\vec{B}^A=0\,,\qquad \partial_t\vec{B}^A=\rot\vec{E}^A\,. \label{b1}
\ee
\item {\bf Definitions}
\be
\vec{B}=\cali \vec{E}\quad\Leftrightarrow\quad \vec{B}_A=G_{AB}\vec{E}^B+\Theta_{AB}\vec{B}^B\,,\quad \vec{E}_A=-G_{AB}\vec{B}^B+\Theta_{AB}\vec{E}^B\,.\label{d1}
\ee
\item \bf{Equations of motion}
\be
dF_A=0\quad\Leftrightarrow\quad \grad\cdot\vec{B}_A=0\,,\qquad \partial_t\vec{B}_A=\rot\vec{E}_A\,.\label{e1}
\ee
\end{itemize}

In description 2 there are $2r$ independent magnetic fields $\vec{B}^a$, furthermore there are as many electric fields $\vec{E}^a$. Note that these last are however not gauge invariant and so not all of their degrees of freedom are physical. The field equations for these electromagnetic fields are the Bianchi identities (\ref{dualbianch}) together with the equations of motion following from varying the Lagrangian (\ref{lag2}) with respect to $\vec{A}^a$:
\begin{itemize}
\item {\bf Bianchi identities}
\be
\grad\cdot\vec{B}^a=0\,,\qquad \partial_t\vec{B}^a=\rot\vec{E}^a\,.\label{b2}
\ee
\item {\bf Equations of motion}
\be
\rot\left(\Omega_{ab}\vec{E}^b+\calg_{ab}\vec{B}^b\right)=0\,.\label{e2}
\ee
\end{itemize}

Let us now show that the equations (\ref{b2}, \ref{e2}) are equivalent to (\ref{b1}, \ref{d1}, \ref{e1}) assuming the identifications (\ref{id12}). Using the relation between the symplectic form $\Omega$, the metric $\calg$ and the complex structure $\cali$ given in (\ref{dualtmetric}) we can solve equation (\ref{e2}) as
\be
\vec{B}^a=\cali^a{}_b\vec{E}^b+\grad \Psi^b\,.\label{selfdeom}
\ee
Due to the gauge freedom (\ref{gaugtransfo0}) we can however always put $\Psi^b=0$. Using the identification (\ref{id12}) that $\vec{B}^a=(\vec{B}_A,\vec{B}^A)$, equation (\ref{e2}) then implies together with the definition (\ref{d1}) that $\vec{E}^a=(\vec{E}_A,\vec{E}^A)$. It then immediately follows that equations (\ref{b2}) are equivalent to equations (\ref{b1}, \ref{e1}). 

\subsection{Scalar field equations}
To compare the scalar field equations of the two theories (\ref{lag1}, \ref{lag2}) it is best to split the problem into two parts. Under the identification (\ref{id12}) we can think of the complex scalars as a function of the real ones: $\phi^A=\phi^A(Y^a)$. As we explicitly checked in section \ref{dualandspecial} we have the following equality:
\be
L_1^{\mathrm{sc. kin.}}(\phi^A(Y^a))=L_2^{\mathrm{sc. kin.}}(Y^a)\,.
\ee
As is well known the Euler-Langrange equations remain satisfied under a scalar field redefinition, i.e.
\be
\partial_\m\frac{\delta L_2^{\mathrm{sc. kin.}}}{\delta \partial_\m Y^a}-\frac{\delta L_2^{\mathrm{sc. kin.}}}{\delta Y^a}=\left(\partial_\mu\frac{\delta L_1^{\mathrm{sc. kin.}}}{\delta\partial_\mu\phi^A}-\frac{\delta L_1^{\mathrm{sc. kin.}}}{\partial \phi^A}\right)\frac{\partial \phi^A}{\delta Y^a}+\mathrm{c.c.}\,.\label{kinredef}
\ee

It is important to realize that the same reasoning doesn't work directly for the other terms in the Lagrangians. The variation principle for the scalar field equations in description 1 is to vary the scalars $\phi^A$ while holding $\vec{E}^A$ and $\vec{B}^{A}$ fixed, while in description 2 we vary $Y^a$ while keeping $\vec{B}^a=(\vec{B}_A,\vec{B}^A)$ fixed. Since the relation $\vec{B}_A=G_{AB}\vec{E}^B+\Theta_{AB}\vec{B}^B$ depends explicitly on the scalars we cannot simply relate the parts of the Lagrangians involving electromagnetic fields by a field redefinition while preserving the variation principle. However, under such a naive field redefinition using the identifications (\ref{id12}) the two Lagrangians (\ref{lag1}, \ref{lag2}) are actually not equal. We will show below that the difference exactly compensates for the changed variation principle and that the two effects nicely cancel out, so that
\be
\frac{\delta L_2^{\mathrm{m.}+\mathrm{sc.}}}{\delta Y^a}=\left(\frac{\delta L_2^{\mathrm{e.m.}}}{\delta\phi^A}\right)\frac{\partial\phi^A}{\partial Y^a}+\mathrm{c.c.}\,.\label{emterms}
\ee
This relation then combines with (\ref{kinredef}) to yield
\be
\partial_\m\frac{\delta L_2}{\delta \partial_\m Y^a}-\frac{\delta L_2}{\delta Y^a}=\left(\partial_\mu\frac{\delta L_1}{\delta\partial_\mu\phi^A}-\frac{\delta L_1}{\delta \phi^A}\right)\frac{\partial \phi^A}{\partial Y^a}+\mathrm{c.c.}\,.
\ee
Since $\phi^A(Y^a)$ is by construction assumed to be invertible we thus find that also the scalar equations of motion of the two descriptions are equivalent.

To show that (\ref{emterms}) holds, let us start in description 1. Varying $L_1^{\mathrm{e.m.}}$ with respect to $\phi^A$ while keeping $\vec{B}^A$ and $\vec{E}^A$ fixed one finds
\be
\frac{\delta L_1^{\mathrm{e.m.}}}{\delta\phi^A}=\frac{1}{2}\left(\partial_AG_{BC}\right)(\vec{B}^B\cdot\vec{B}^C-\vec{E}^B\cdot\vec{E}^C)-\left(\partial_A\Theta_{BC}\right)\vec{E}^B\cdot\vec{B}^C\,.\label{sceq1}
\ee 
Similarly we can compute the variation of $L_2^{\mathrm{m.}+\mathrm{sc.}}$ with respect to $Y^a$ while holding $\vec{B}^a=(\vec{B}_A,\vec{B}^A)$ fixed, the result is simply
\bea
\frac{\delta L_2^{\mathrm{m.}+\mathrm{sc.}}}{\delta Y^a}&=&\frac{1}{2}\left(\partial_a\calg_{bc}\right)\vec{B}^b\vec{B}^c\,. \label{effe}
\eea
Now that we have completed the variation principle we are free to use the identifications (\ref{id12}). After some careful algebra it follows that
\be
\frac{1}{2}\left(\partial_a\calg_{bc}\right)\vec{B}^b\vec{B}^c=\frac{1}{2}\left(\partial_aG_{AB}\right)\left(\vec{B}^A\cdot\vec{B}^B-\vec{E}^A\cdot\vec{E}^B\right)-\left(\partial_a\Theta_{AB}\right)\vec{E}^A\cdot\vec{B}^B\,.\label{sceq2}
\ee
To see that (\ref{emterms}) is indeed correct it is now enough to use the fact that for any real function $f(\phi,\bar\phi)$ the 'chain rule' reads $\partial_a f(\phi(Y),\bar\phi(Y))=\frac{\partial f}{\partial \phi^A}\partial_a\phi^A+\frac{\partial f}{\partial \bar\phi^A}\partial_a\bar\phi^A$, and compare (\ref{effe}, \ref{sceq2}) to (\ref{sceq1}) and it's complex conjugate.

\bibliographystyle{JHEP}
\bibliography{n2}

\end{document}